\documentclass[preprint,amsmath,amssymb,pre]{revtex4-1}
\usepackage{color,longtable}
\usepackage{graphicx}
\usepackage{amssymb,amsfonts,amsmath}
\begin{document}%
\title{Bond-number-controlled durability of cohesive granular materials under repeated vibration}
\author{Hikari Yokota}
\author{Rei Kurita}

\affiliation{%
Department of Physics, Tokyo Metropolitan University, 1-1 Minamioosawa, Hachiouji-shi, Tokyo 192-0397, Japan
}%
\date{\today}

\begin{abstract}
Cohesive granular materials derive their mechanical stability not only from the strength of individual interparticle bonds but also from the number of bonds forming the load-bearing network. However, these two effects are difficult to separate experimentally because conventional control parameters, such as liquid content, generally alter both simultaneously. Here, we use a mixed granular system composed of cohesive and noncohesive grains to control the cohesive bond number while keeping the bond strength approximately unchanged. We investigate the failure lifetime under repeated vibration and find that the number of cycles to failure, $N_f$, depends strongly on the mixing ratio $\alpha$. In a mean-field picture of random mixing, the fraction of cohesive contacts scales as $\alpha^2$, and $N_f$ increases approximately exponentially with $\alpha^2$.
By contrast, although the lifetime tends to decrease with increasing vibration intensity $G$, its dependence on $G$ is comparatively weak over the present experimental range. Remarkably, although the Young's modulus is nearly independent of $\alpha$ above the rigidity threshold, the lifetime continues to increase strongly with $\alpha$. This demonstrates that mechanical rigidity and durability against repeated perturbations exhibit distinct dependences on the cohesive network. These results identify bond number as a key control parameter for the durability of cohesive granular materials.
\end{abstract}

\maketitle

\section{Introduction}
The mechanical properties of granular materials vary greatly with the amount and state of the liquid they contain. From dry granular materials to partially wet granular materials containing a small amount of liquid, and further to slurries and fully immersed granular materials, both interparticle interactions and macroscopic mechanical responses change markedly with liquid content~\cite{mitarai2006,samadani2000,scheel2008}. In partially wet granular materials, in particular, a small amount of liquid or surface coating can form capillary or cohesive bridges between particles, giving rise to shape retention, yield stress, cohesion, and rigidity. Such materials are encountered in a wide range of everyday and industrial applications, including construction materials, asphalt, porous materials, and pharmaceutical tablets~ \cite{duran2012,brown2016,nagel1992,jaeger1992,guyon1994}.

At the microscopic level, the mechanical response of partially wet granular materials is governed by two distinct factors: the strength of individual interparticle bonds and the number and network structure of those bonds. 
For systems in which cohesive bonds are formed relatively uniformly throughout the material, macroscopic elastic properties can often be described well in terms of the strength of individual interparticle bonds~\cite{Zaccone2007,Bonn2007,Bonn2012}. 
In many practical granular materials, however, particle size, wettability, shape, and local contact conditions are spatially heterogeneous, leading to variations in both bond strength and bond number. 
For example, in cementitious materials, a broad particle-size distribution leads to nonuniform hydration, and macroscopic hardening develops as the hydrated regions form a percolated network~\cite{Scherer2012}. 
In such heterogeneous systems, the number and spatial arrangement of cohesive contacts can therefore play an important role in addition to the local bond strength. 
Because the mechanical properties and functionality of mixed granular materials are often controlled through composition, clarifying the role of bond number and network connectivity is also important from the viewpoint of material design.

To overcome this difficulty, we have previously employed a model granular system consisting of a mixture of coated grains and bare sand~\cite{Tani2021, Fujio2024}. In this system, cohesive bonds are formed predominantly through contacts between coated grains, allowing the average bond number to be controlled by the mixing fraction $\alpha$ without substantially changing the strength of individual bonds. Previous studies have shown that a system-spanning cohesive network emerges through connective percolation at around $\alpha \sim 0.23$~\cite{Tani2021}. This is close to the connectivity percolation value of $\alpha \sim 0.25$~\cite{stauffer1994}. Meanwhile, rigidity percolation, associated with the formation of a mechanically rigid network, occurs at $\alpha \gtrsim 0.6$~\cite{Fujio2024}. Interestingly, the Young's modulus becomes nearly independent of $\alpha$ above this rigidity threshold. This indicates that once a rigid network is established, further increasing the bond number has little effect on the small-strain stiffness. It also suggests that a certain fraction of noncohesive or functional particles may be incorporated without substantially compromising the rigidity of the material.

Previous studies have mainly focused on static or quasistatic mechanical properties under small deformation. In practical materials, however, durability against repeated mechanical perturbations is also an important property. High stiffness under small deformation does not necessarily imply a long lifetime under cyclic loading. In particular, although the Young's modulus saturates above the rigidity-percolation threshold, additional cohesive bonds within the network may still enhance resistance to repeated perturbations. In this study, we therefore use the mixed granular system described above to investigate how the failure lifetime under repeated vibration depends on the bond number. We show that the failure lifetime exhibits a strong nonlinear dependence on the cohesive-contact fraction, whereas the dependence on vibration intensity is comparatively weak over the present experimental range.
This contrast highlights the distinct roles of bond number in static rigidity and durability under repeated perturbations.

\section*{Results and Discussion}
\begin{figure}[htbp]
\begin{center}
\includegraphics[width=120mm]{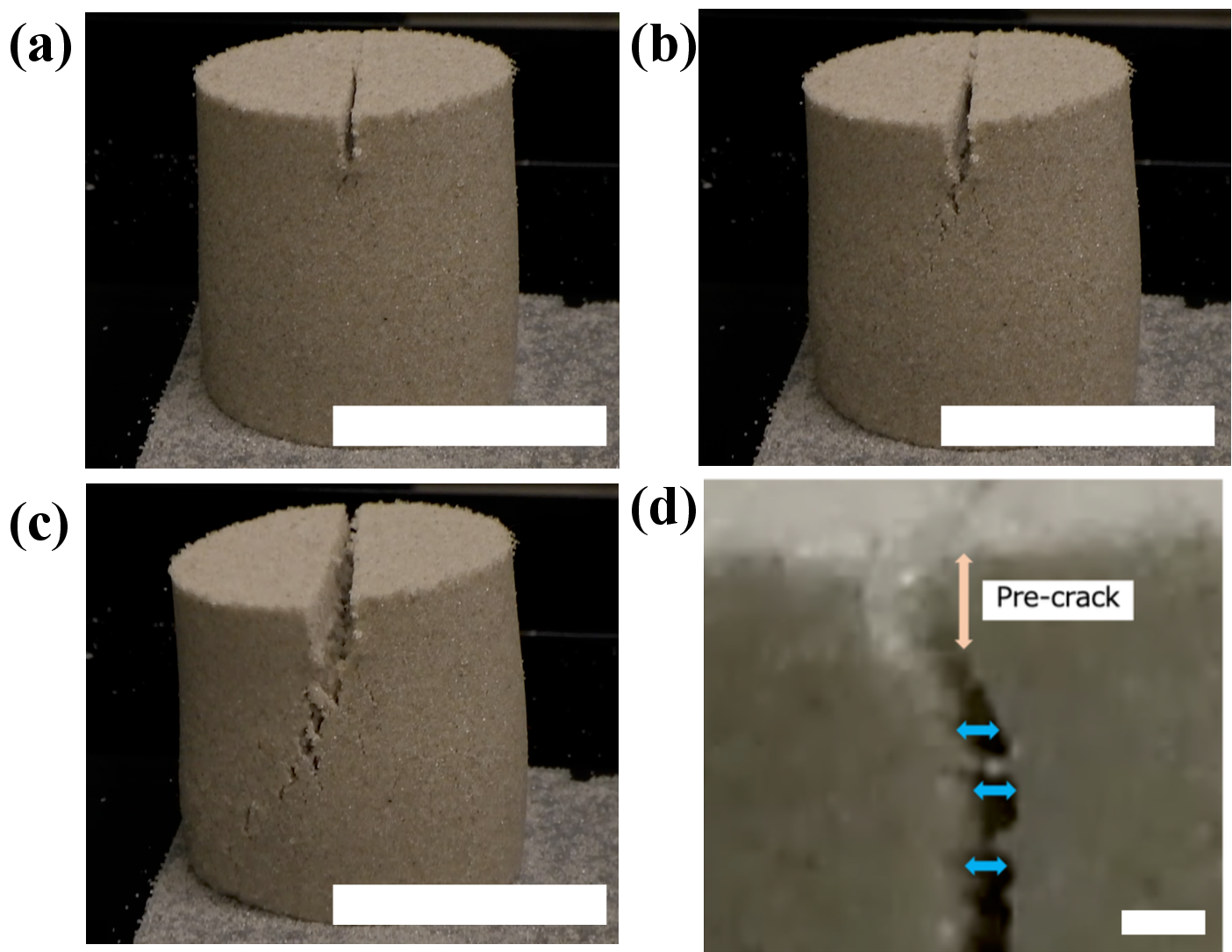}
\end{center}
\caption{
Time evolution of a granular block under repeated vibration at 40~Hz and $\alpha=0.6$.
(a) At $t = 80$~s, the sample remains almost unchanged from its initial state.
(b) At $t = 90$~s, a crack begins to develop from the pre-crack.
(c) At $t = 95$~s, the crack has rapidly propagated and the sample is close to complete collapse.
The white scale bars in (a)--(c) represent 50~mm.
(d) Enlarged view of a crack in another sample under the same conditions (40~Hz and $\alpha=0.6$) at $t=100$~s.
The crack propagates from the pre-crack.
The white scale bar in (d) represents 3~mm.
The crack width was measured at three positions, and the average value was defined as $w$.
}
\label{image}
\end{figure}

Kinetic sand was mixed with noncohesive sand at a mass fraction $\alpha$, and the mixture was compacted into a cylindrical granular block in a random close-packed (RCP) state. A pre-crack was introduced at the surface of the block before repeated vibration was applied (see Methods). Cyclic loading is widely used to characterize the failure lifetime of materials~\cite{Dowling1996}.
Figure~\ref{image}(a)--(c) shows the failure process of a granular block under repeated vibration at $\alpha$ = 0.6. No appreciable macroscopic change was observed for some time after the onset of vibration. Beyond a certain time, however, a crack rapidly developed from the vicinity of the pre-crack and propagated through the sample, eventually leading to collapse of the entire block. Figure~\ref{image}(d) shows an enlarged view of the crack. At each time, the crack width was measured at approximately three positions, and their average was defined as $w$.

\begin{figure}[htbp]
\begin{center}
\includegraphics[width=120mm]{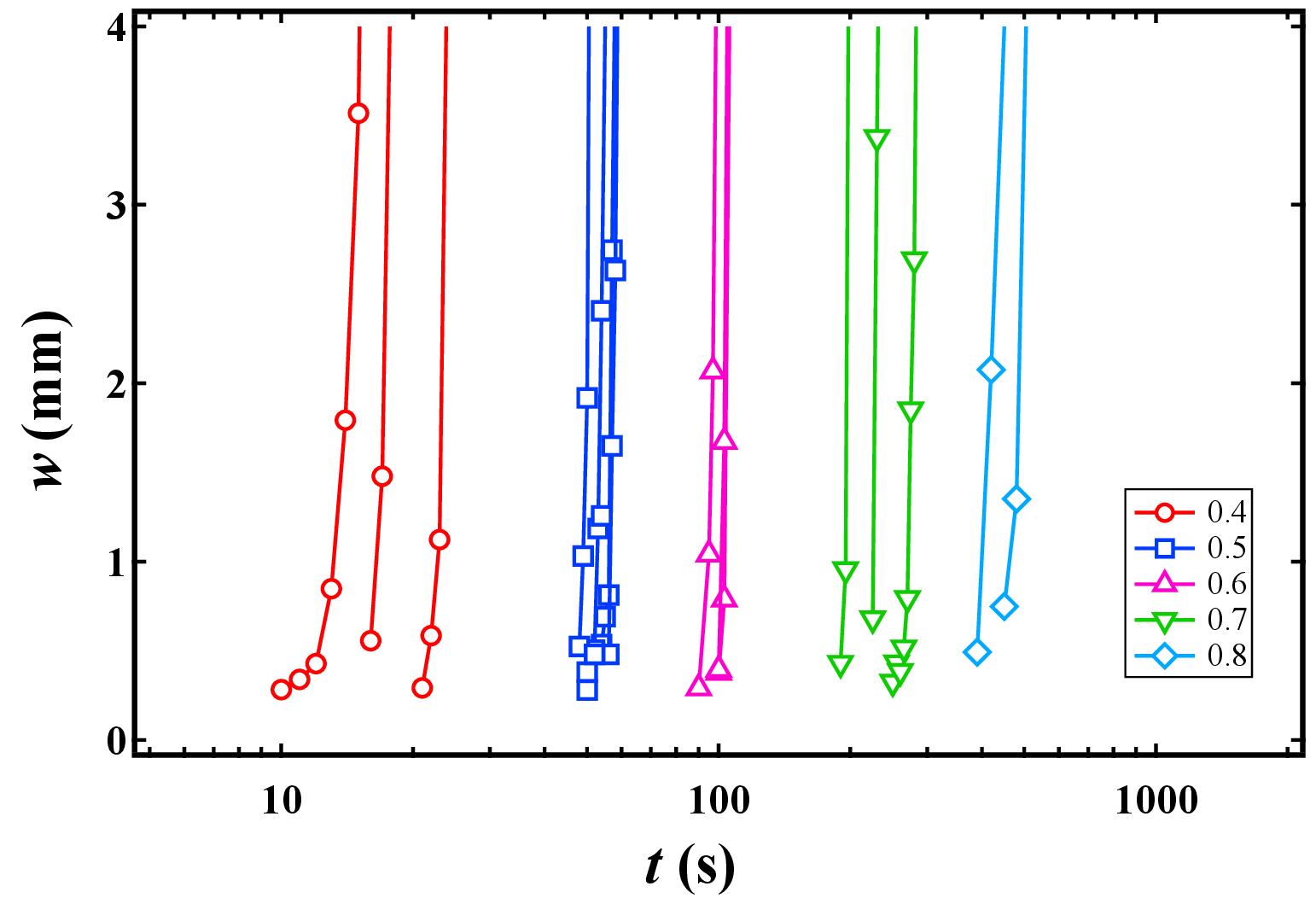}
\end{center}
\caption{
Time evolution of the crack width $w$ for granular blocks with different kinetic-sand fractions $\alpha=0.4$--0.8 under repeated vibration.
The crack width was obtained by averaging measurements at approximately three positions along the crack.
Data are shown only after a visible crack appeared; the period before crack initiation is therefore not plotted.
For all $\alpha$, $w$ increases rapidly after crack initiation, and the onset of crack growth shifts to longer times as $\alpha$ increases.
}
\label{width}
\end{figure}

Figure~\ref{width} shows representative time evolutions of $w$ for granular blocks with $\alpha$ = 0.4--0.8. The vibration frequency was fixed at 50~Hz and the amplitude at 0.31~mm. 
No measurable crack was observed during the initial stage, and these periods are not included in Fig.~\ref{width}.
Once a visible crack appeared, $w$ increased rapidly.
The onset of crack growth shifted to later times with increasing $\alpha$.
To quantify the failure process, we defined failure as the point at which $w$ reached 2.0~mm. 
The number of vibration cycles applied up to this point was defined as the failure lifetime, $N_f$.

\begin{figure}[htbp]
\begin{center}
\includegraphics[width=120mm]{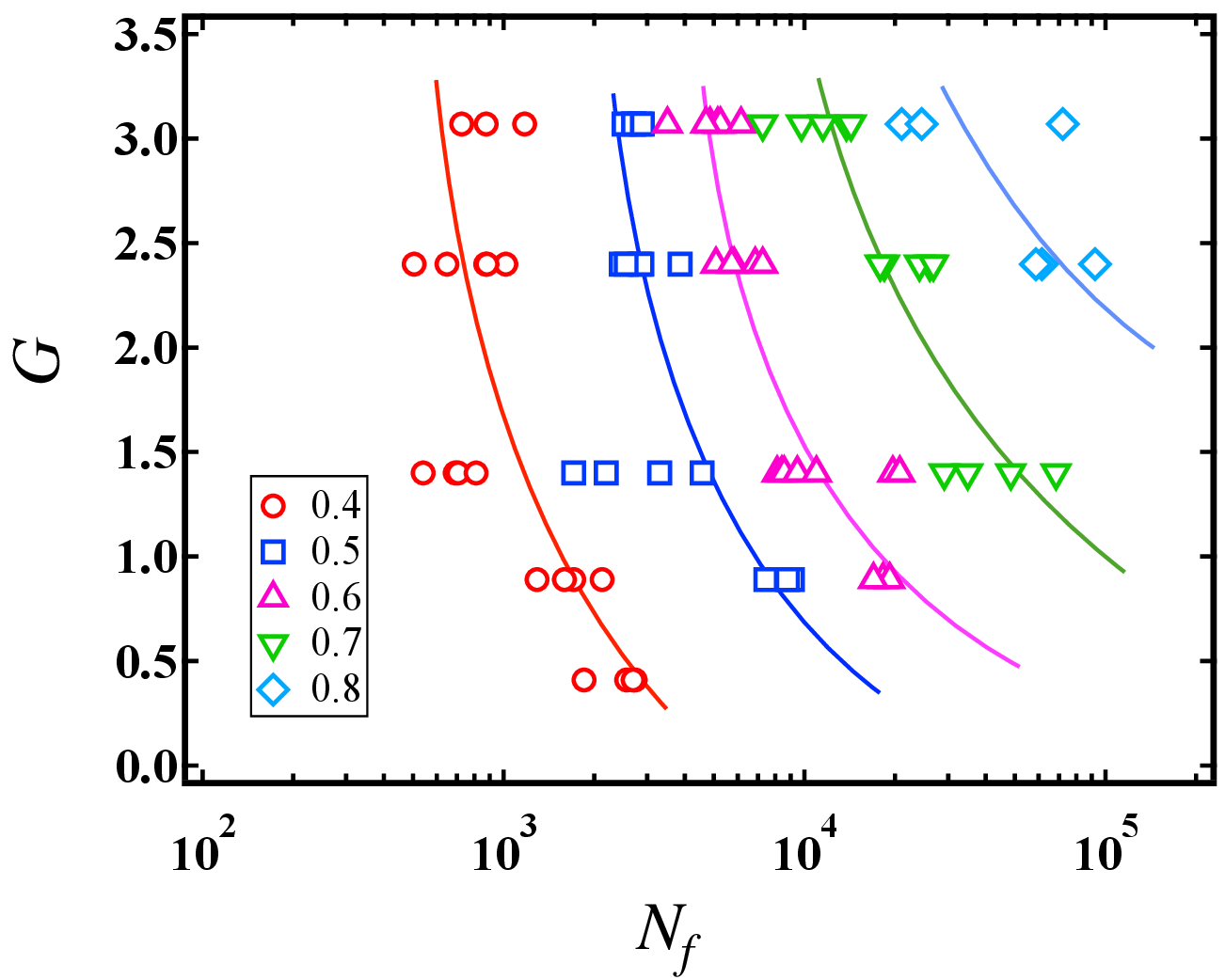}
\end{center}
\caption{
Relation between the dimensionless peak acceleration $G$ and the failure lifetime $N_f$ for granular blocks with different kinetic-sand fractions $\alpha=0.4$--0.8.
The pre-crack depth was fixed at 9~mm.
With increasing $\alpha$, the $G$--$N_f$ relation systematically shifts toward longer lifetimes, indicating enhanced resistance to repeated vibration.
For each $\alpha$, the failure lifetime generally increases as $G$ decreases.
The solid lines are guides to the eye.
}
\label{lifetime}
\end{figure}

To evaluate the stability against repeated vibration, we examined the relation between the number of cycles to failure $N_f$ and the peak acceleration normalized by gravitational acceleration, $G$. 
This representation is analogous to the conventional S--N curve widely used in fatigue tests, with $G$ corresponding to the intensity of cyclic loading~\cite{Dowling1996}.
Figure~\ref{lifetime} shows the relation between $G$ and $N_f$ for different values of $\alpha$. For each $\alpha$, $N_f$ generally increases as $G$ decreases. Moreover, the $G$--$N_f$ relation shifts toward longer lifetimes with increasing $\alpha$, demonstrating that increasing the fraction of cohesive grains strongly enhances the resistance to repeated vibration.

For $\alpha \geq 0.9$, the experimentally accessible range was limited. In this regime, failure was sometimes observed at larger $G$, whereas no failure occurred within the 1~h observation window under weaker vibration. However, the data in this range showed relatively large scatter, and at large $G$ possible heating effects of the vibration apparatus could not be completely excluded. We therefore restrict the following quantitative analysis mainly to $\alpha \leq 0.8$.

The strong dependence of the lifetime on $\alpha$ suggests that the number of cohesive bonds plays a central role in the resistance to repeated vibration.
Within a mean-field picture assuming random mixing of coated and uncoated grains, the probability that both grains forming a contact are coated is proportional to $\alpha^2$.
Thus, $\alpha^2$ provides a simple mean-field estimate of the fraction of cohesive contacts in the granular network.

\begin{figure}[htbp]
\begin{center}
\includegraphics[width=120mm]{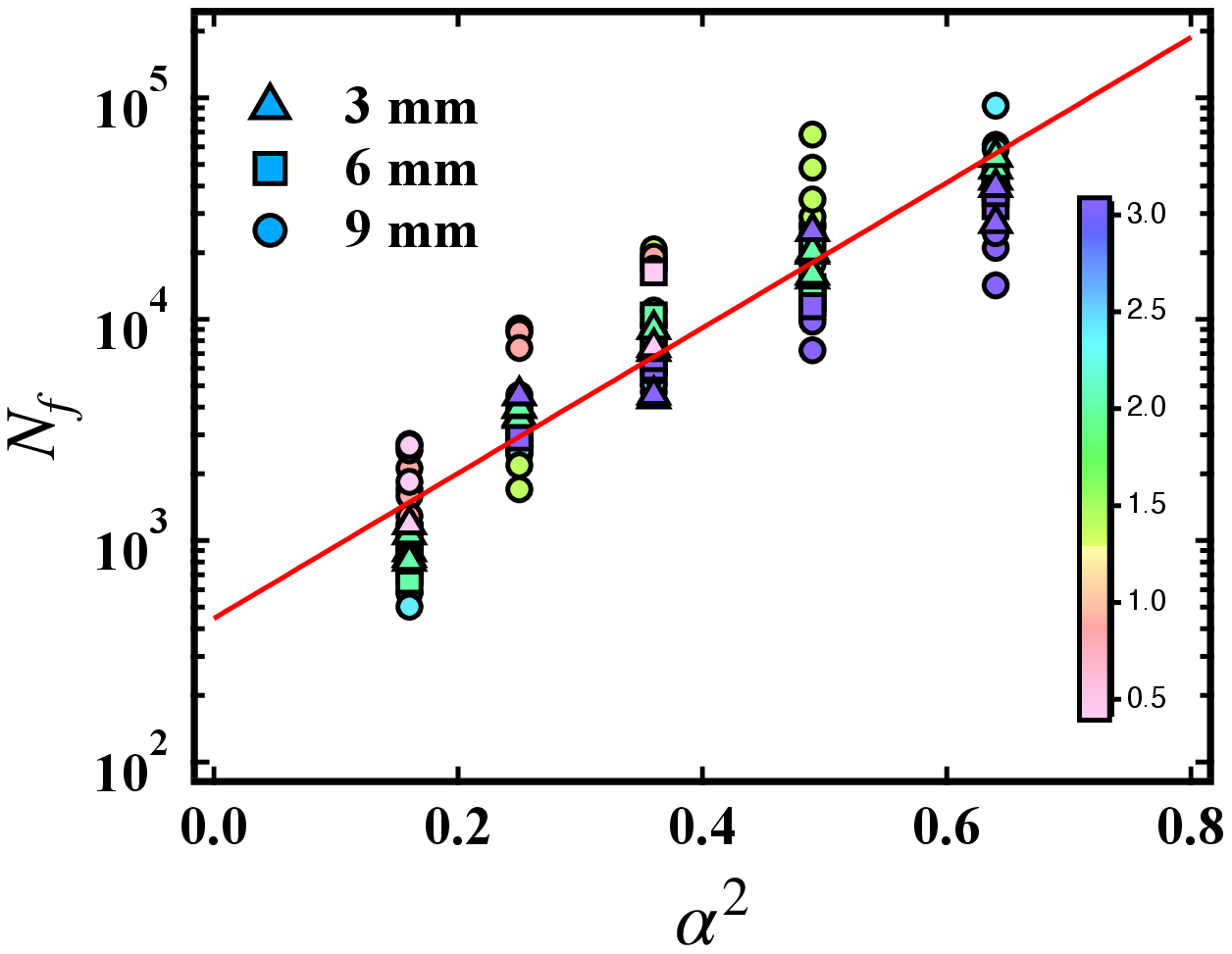}
\end{center}
\caption{
Failure lifetime $N_f$ as a function of the mean-field cohesive-contact fraction $\alpha^2$.
The color indicates the vibration intensity $G$, while triangles, squares, and circles correspond to pre-crack depths of 3, 6, and 9~mm, respectively.
The failure lifetime increases strongly and nonlinearly with $\alpha^2$ despite variations in $G$ and pre-crack depth.
The solid line represents the phenomenological fit given by Eq.~\ref{eq:Nf}.
}
\label{scaling}
\end{figure}

Figure~\ref{scaling} shows $N_f$ as a function of $\alpha^2$.
The colors indicate the vibration intensity $G$, while triangles, squares, and circles correspond to pre-crack depths of 3, 6, and 9~mm, respectively.
Despite variations in $G$ and pre-crack depth, the data exhibit a strong nonlinear increase of $N_f$ with $\alpha^2$.
As a phenomenological description, the dependence is well represented by
\begin{equation}
N_f = N_0 \exp\left(C\alpha^2\right),
\label{eq:Nf}
\end{equation}
with $N_0 = 446$ and $C = 7.6$.

The dependence on vibration intensity is comparatively weak within the present experimental range.
At a given $\alpha$, larger $G$ generally tends to reduce the lifetime, but the present data do not establish a specific functional dependence on $G$.
These results indicate that the cohesive-bond fraction is the dominant control parameter for the failure lifetime under the conditions examined here.

The weak dependence of $N_f$ on the pre-crack depth indicates that the lifetime is not primarily controlled by the growth of a pre-existing macroscopic crack.
Instead, the strong dependence of $N_f$ on $\alpha$ suggests that the cohesive-bond network plays a dominant role in determining the durability under repeated vibration.
This is consistent with the observation that the macroscopic crack develops rapidly only near failure, whereas the lifetime itself varies strongly with the cohesive-grain fraction.

A particularly important feature of the present results is that the lifetime remains strongly dependent on $\alpha$ even for $\alpha \geq 0.6$, where the Young's modulus has already reached an approximately constant value.
Thus, the formation of a mechanically rigid network is sufficient to determine the small-strain stiffness, whereas resistance to repeated perturbations remains sensitive to the number of cohesive bonds within that network.
In other words, additional cohesive bonds beyond the rigidity threshold do not significantly increase the elastic stiffness, but substantially enhance the durability of the granular material.

The approximately exponential increase of $N_f$ with $\alpha^2$ further shows that the effect of the cohesive-bond number on durability is strongly nonlinear.
This strong nonlinearity cannot be understood simply as an additive effect in which a larger initial number of cohesive bonds merely increases the number of bonds that must be lost before failure.
For example, if cohesive bonds rupture independently with a constant probability $q$ per vibration cycle and failure occurs when the remaining cohesive-bond fraction reaches a fixed threshold $\alpha_c^2$, then
\[
\alpha_c^2=\alpha^2(1-q)^{N_f},
\]
which gives
\[
N_f \sim \ln \alpha
\]
for $q \ll 1$.
This dependence is much weaker than that observed experimentally.
Thus, although the cohesive-bond number is clearly a major control parameter for durability, the microscopic origin of its strong nonlinear influence remains unresolved.

The dependence on vibration intensity $G$ is comparatively weak over the present experimental range.
Although larger $G$ generally tends to shorten the lifetime, the present data do not establish a specific functional form for this dependence.
Clarifying the microscopic mechanism governing the lifetime, including the origin of the strong nonlinear bond-number dependence and the role of vibration intensity, remains an important subject for future work.

\section*{Summary}
The mechanical properties of cohesive granular materials depend not only on the strength of individual bonds but also on the number of bonds formed within the material. In practical mixed granular materials, the number of cohesive contacts changes with composition, making it important to clarify how bond number governs macroscopic mechanical behavior. Here, we use a mixed granular system composed of cohesive and noncohesive grains to control the bond number while keeping the bond strength approximately unchanged, and investigate the failure lifetime under repeated vibration.

The failure lifetime $N_f$ depends strongly on the mixing fraction $\alpha$. In a mean-field picture of random mixing, the fraction of cohesive contacts is proportional to $\alpha^2$, and $N_f$ increases approximately exponentially with $\alpha^2$. By contrast, although larger vibration intensity $G$ tends to shorten the lifetime, the dependence on $G$ is comparatively weak over the present experimental range. Remarkably, although the Young's modulus nearly saturates for $\alpha \gtrsim 0.6$, $N_f$ continues to increase strongly with $\alpha$. This demonstrates that durability remains sensitive to bond number even after rigidity has been established, and that static stiffness alone does not fully characterize the resistance of the cohesive network to repeated perturbations. In addition, the pre-crack depth has little influence on $N_f$, indicating that the lifetime is not primarily controlled by the growth of a pre-existing macroscopic crack. These results demonstrate that the cohesive-bond number is a major control parameter for durability and that its effect on the failure lifetime is strongly nonlinear.

\section*{Materials and Methods}
For noncohesive grains, we used Tohoku Silica Sand No.~5 (Tohoku Silica Sand Co., Ltd.). The grains had a density of 2.62~g/cm$^3$, with short diameters of 0.08--0.35~mm and long diameters of 0.23--0.55~mm. Kinetic Sand (Rangs Japan), consisting of sand grains coated with silicone oil, was used as the cohesive grains~\cite{Kinetic}. The density of the grains was 2.62~g/cm$^3$, with short diameters of 0.09--0.29~mm and long diameters of 0.16--0.41~mm. 

For each mass fraction $\alpha$ of kinetic sand, the required amounts of kinetic sand and noncohesive sand were weighed, spread on the same metal tray, and thoroughly mixed by hand. The spatial homogeneity of the mixture was evaluated using blue-colored kinetic sand. The degree of inhomogeneity was estimated from the width of the spatial distribution of the blue intensity and was less than 1.6 \% for all values of $\alpha$. Details of this procedure are described in our previous study~\cite{Fujio2024}. Since this variation is comparable to the experimental uncertainty, the mixtures were regarded as spatially homogeneous.

Random close-packed (RCP) granular blocks were prepared as follows. The mixed grains were gently poured into a cylindrical container with a diameter of 56.0~mm and a height of 63.0~mm at a rate of approximately 10~g/s. The grains were then compacted by applying an external pressure of approximately $5 \times 10^4$~Pa using a plate whose area was nearly identical to the cross-sectional area of the container. Additional grains were subsequently added, followed by further compaction. This procedure was repeated approximately 25 times until the container was filled. The packing fraction of the resulting granular blocks was $61 \pm 2$\%~\cite{Tani2021}, corresponding approximately to the RCP state.
A pre-crack was introduced at the center of the upper surface of the granular block using a 1-mm-diameter wire. To ensure that the pre-crack was introduced at the same position for every sample, a narrow guide slit was made at the center of the acrylic plate used for compaction during preparation of the RCP block. The wire was inserted through this guide slit to a prescribed depth and then carefully removed before vibration was applied.
The pre-crack depth was varied between 3, 6, and 9~mm.

Sinusoidal vibration was generated using a vibration exciter (SVA, San-esu Co., Ltd.) connected to a function generator (AFG1062, Tektronix). An acrylic tray was horizontally fixed to the vibration exciter with screws, and the granular block was placed on the tray. To prevent the sample from detaching from the tray during vibration, double-sided adhesive tape was placed beneath the sample.
The vibration frequency $f$ was varied between 30 and 50~Hz. At frequencies of 60~Hz or higher, the granular block detached from the adhesive tape even at the minimum attainable vibration amplitude, and these conditions were therefore excluded from the measurements. The actual motion of the acrylic tray was monitored using a laser displacement sensor (CDX, Optex FA Co., Ltd.). The displacement was confirmed to be approximately sinusoidal, and the vibration amplitude $A$ was determined directly from the measured displacement. The dimensionless peak acceleration $G$ was calculated as
$G = A(2\pi f)^2/g$,  where $g$ is the gravitational acceleration.
The deformation and failure of the granular blocks were recorded using a digital camera (IXY650, Canon), and the acquired images were analyzed using ImageJ.
Images were recorded at intervals of 1~s.

\section*{Acknowledgments}
R. K. was supported by JSPS KAKENHI Grant Number 26K00676. 

\section*{Data availability}
All data supporting the findings of this study are included within the article.

\section*{AUTHORS CONTRIBUTIONS}
R.~K. conceived the project. H.~Y performed the experiments. H. Y. and R. K. analyzed the data. R.~K. wrote the manuscript.

\section*{COMPETING INTERESTS STATEMENT}
The authors declare that they have no competing interests. 

\section*{CORRESPONDENCE}
Correspondence and requests for materials should be addressed to R.~K. (kurita@tmu.ac.jp).


\clearpage

\end{document}